\documentclass[letterpaper,preprintnumbers,amsmath,amssymb,nofootinbib,onecolumn,superscriptaddress,notitlepage]{revtex4}
\pdfoutput=1
\usepackage{graphicx}
\usepackage{dcolumn}
\usepackage{bm}
\usepackage[usenames,dvipsnames]{color}
\usepackage{amsmath}	
\usepackage{amssymb}	
\usepackage{multirow}
\usepackage{gensymb}
\usepackage{xspace}
\usepackage{upgreek}
\usepackage[colorlinks=true,citecolor=blue,linkcolor=blue]{hyperref}

\newcommand{\nufate}{{$\nu$}FATE\xspace}

\begin{document}

\title{High-energy neutrino attenuation in the Earth and its associated uncertainties}

\author{Aaron C. Vincent}
\email{aaron.vincent@queensu.ca}
\affiliation{Department of Physics, Imperial College London, London SW7 2AZ, UK}
\affiliation{Department of Physics, Engineering Physics and Astronomy, Queen's University, Kingston ON K7L 3N6, Canada}
\author{Carlos A. Arg\"uelles}
\email{caad@mit.edu}
\affiliation{Department of Physics, Massachusetts Institute of Technology, Cambridge, MA 02139, USA}
\author{Ali Kheirandish}%
 \email{ali.kheirandish@icecube.wisc.edu}
\affiliation{Department of Physics and Wisconsin IceCube Particle Astrophysics Center, University of Wisconsin, Madison, WI 53706, USA}

\begin{abstract}
We describe \nufate: Neutrino Fast Attenuation Through Earth, a very rapid method of accurately computing the attenuation of high-energy neutrinos during their passage through Earth to detectors such as IceCube, ANTARES or KM3Net, including production of secondary neutrinos from $\tau^\pm$ lepton decay. We then use this method to quantify the error on attenuation due to uncertainties in the isotropic neutrino spectrum, the composition of the Earth, and the parton distribution functions. We show that these can be as large as 20\%, which can significantly impact reconstructed astrophysical neutrino parameters, as well as searches for new physics. An implementation of this algorithm is provided as a public code.\footnote{\url{https://github.com/aaronvincent/nuFATE}}
\end{abstract}
\maketitle

\section{Introduction}
It is well known that the neutrino-nucleus scattering cross section grows with energy, and at lab frame energies above a TeV the mean free path of neutrinos in the Earth falls below the distance traveled: $(\sigma \rho_\oplus/m_p)^{-1} < R_\oplus$. With the advent of the first extraterrestrial events seen by IceCube, via both contained vertex high-energy starting events (HESE) and thoroughgoing muon tracks from the Northern Hemisphere, the practicalities of neutrino detection become important to theorists and observers alike.

As more data become available, and as analyses of high-energy neutrino detection become more sophisticated, the rapid and accurate computation of neutrino attenuation becomes essential. Although this effect is accounted for by the effective areas provided by IceCube \cite{Aartsen:2013jdh, Aartsen:2014gkd, Aartsen:2015rwa}, it must be explicitly recomputed for analyses that rely on different hypotheses. These include any exploration of uncertainties in standard scenarios: different parton distribution function (PDF) sets, different models of the Earth,  uncertainties in the cosmic neutrinos' spectral index, and any new physics that affects the neutrino flux  to some degree before or during propagation through the Earth to the detector. Ordinary ODE methods of solving the propagation equations, such as the ones implemented in the publicly available nuSQuIDS package \cite{squids,nusquids}, are computationally quite expensive, since they require the solution for a system of coupled equations at each zenith angle. This can require numerical integration over two to three variables at each step, thus limiting the ability to rapidly scan over the space of parameters that may affect attenuation.  

Here, we present a simple method of solving the attenuation equations numerically, which relies on basic linear algebra. Once a solution is found via an eigenvalue decomposition, it can be evaluated for any neutrino energy and optical depth at no extra cost. By separating out the particle and propagation components, numerical integration only needs to be done once per particle physics (and PDF) model, leading to faster computation of attenuation rates by orders of magnitude. This work is structured as follows: in Section \ref{sec:eqs} we remind the reader of the cascade equations that govern neutrino attenuation and regeneration. Section \ref{sec:methods} explains the \nufate solution method, and Section \ref{sec:examples} shows powerful examples of its application: these include 1) systematic effects on effective area induced by uncertainties in the astrophysical flux; 2) an exploration of uncertainties due to variation of the reference Earth model; and 3) The effect of using different PDF sets and their associated errors on the attenuated flux. Finally, Appendix \ref{sec:app} explains how to run and modify the publicly available code. 

\section{Cascade equations}
\label{sec:eqs}
The attenuation of neutrinos of flavor $\ell = \{\nu_e,\nu_\mu,\nu_\tau,\bar \nu_e,\bar\nu_\mu,\bar\nu_\tau\}$ during their passage through Earth can be described by the cascade equation \cite{Nicolaidis:1996qu, Naumov:1998sf, Kwiecinski:1998yf, Iyer:1999wu, Giesel:2003hj, Rakshit:2006yi,Palomares-Ruiz:2015mka}
\begin{eqnarray}
\frac{\partial}{\partial x} \left(\frac{d\phi_{\nu_\ell} (E_\nu, x)}{dE_\nu}\right)  =  -  \, \left(\sigma^{\textrm{NC}}_{\nu_\ell} (E_\nu) + \sigma^{\textrm{CC}}_{\nu_\ell} (E_\nu) \right) \, \frac{d\phi_{\nu_\ell}(E_\nu, x)}{dE_\nu}  
+ \, \int_{E}^{\infty} \, d \tilde E \, \frac{d\sigma^\textrm{NC}_{\nu_\ell} (E_\nu,\tilde E_\nu)}{d E_\nu} \, \frac{d\phi_{\nu_\ell} (\tilde E_\nu, x)}{d\tilde E_\nu} ~.
\label{eq:transpN}
\end{eqnarray}

The target column density $x$ depends only on the zenith angle $\theta$ and is the integral along the line of sight of the Earth density:
\begin{equation}
x(\theta) \equiv N_a \int_{l.o.s.} \rho_\oplus[r(x,\theta)] dx.
\label{eq:coldens}
\end{equation}
The cross sections in the first term on the right of Eq. \eqref{eq:transpN} are the total charged-current (CC) and neutral current (NC) neutrino-nucleus cross sections. The last term comes from scattering of neutrinos of energy $\tilde E$ to $E$ via a neutral-current interaction. Each cross section furthermore includes the scattering rate with protons and neutrons in proportion with their abundance. Taking the Earth as isoscalar, this reduces to an average of the $\nu-p$ and $\nu-n$ cross sections. In the case of electron antineutrinos, the important Glashow resonance at $E_\nu \sim 6.3$ PeV becomes non-negligible. The antineutrino-electron cross section must therefore be added to both terms in Eq. \eqref{eq:transpN}.

For tau (and antitau) neutrinos, a third term contributes to Eq. \eqref{eq:coldens}, to account for CC production of tau (or antitau) leptons, which then decay back to tau neutrinos \cite{Halzen:1998be,Iyer:1999wu,Dutta:2000jv,Hettlage:2001yf,Tseng:2003pn,Hussain:2003vi,Yoshida:2003js,2004ApJ...613.1285F,Reya:2005vh}. This regeneration effect is often parametrized by simultaneously solving the $\nu_\tau$ and $\tau^-$ (or $\bar \nu_\tau$ and $\tau^+$) cascade equations. However, for energies below a few PeV, the tau lepton does not live very long with respect to the propagation distance (meters versus thousands of km), so it is sufficient to treat $\nu_\tau$ regeneration ``on-the-spot'' (see \cite{Fedynitch:2015zma} for a similar approach). One adds the following term to Eq. \eqref{eq:transpN}
\begin{equation}
I_\tau = \int_E^\infty {d \tilde E} \int_0^1 dy \frac{z}{E} \frac{dn}{dz} \frac{d\sigma^{CC}_{\nu_\tau}(\tilde E,y)}{dy} \frac{d\phi_{\nu_\tau}(\tilde E,x)}{d \tilde E}. \label{eq:taudec}
\end{equation}
Once again, this represents the rate of obtaining a neutrino of energy $E$ from any higher energy $\tilde E$. The energy of the intermediate tau particle is $E_\tau =  \tilde E(1-y)$, and $z = E/E_\tau = E/\tilde E (1-y)$. The spectrum $dn/dz$ is the decay distribution of tau neutrinos from tau decay. A parametrization of $dn/dz$ for all decay channels can be found in Appendix A of \cite{Dutta:2000jv}. 

For very hard spectra, decaying tau leptons can also contribute to the $\nu_e$ and $\nu_\mu$  (and $\bar \nu_e$ and $\bar \nu_\mu$) fluxes \cite{Beacom:2001xn}. In this case, the $\nu_{\ell = \{e,\mu\}}$ cascade equations must be solved simultaneously to the $\nu_\tau$ equation, adding a term identical to Eq. \eqref{eq:taudec}, but with $dn/dz$ replaced with the spectrum of daughter electron or muon neutrinos\footnote{As in Ref. \cite{Palomares-Ruiz:2015mka}, we use
\begin{equation}
\frac{d n_{\tau\rightarrow \nu_{\ell = \{e,\mu\}}}}{dz} \simeq 0.18(4 - 12z +12z^2 - 4 z^3).
\end{equation}}. This has the effect of mixing the flavour ODEs. 

\section{Fast solution}
\label{sec:methods}
The method used here was developed for Ref. \cite{Arguelles:2017atb} (see also \cite{Arguelles:2016zvm}, and a similar approach used in Refs. \cite{1993Lipari,Fedynitch:2015zma}) to solve the cascade equation that occurs in hypothetical dark matter-neutrino scattering in the Galactic halo, and adapted with an eye towards upcoming studies of new physics at IceCube. The equations (\ref{eq:transpN}, \ref{eq:taudec}) can be broadcast into a specific shape. Using the shorthand $\varphi(E) \equiv d\phi_{\nu_\ell} /d E$,:
\begin{equation}
\frac{d\varphi(E)}{dx} = -\sigma(E)\varphi(E) + \int_E^\infty d\tilde E f(E,\tilde E) \varphi(\tilde E)
\label{genericeq}
\end{equation}
This can be solved with the standard numerical procedure of splitting the energy range into a ``vector'' of N elements: $E_\nu \rightarrow \vec E_\nu$, $\varphi\rightarrow \vec \phi$. 
 Then
\begin{equation} 
\frac{d\vec \phi}{dx} = (- \mathrm{diag}(\vec \sigma) + C) \vec \phi = M \vec \phi. \label{vectoreq}
\end{equation}
The components of the $N\times N$ matrix $C$ are the integrand in Eq. \eqref{genericeq}: $C_{ij} = f(E_i,\tilde E_j)$. Since Eq. \eqref{vectoreq} is linear, the eigenvectors $\hat \phi_i$ of $M$ individually satisfy the differential equation $\hat \phi_i' = \lambda_i \hat \phi_i$, where $\lambda_i$ are the corresponding eigenvalues. This is the key to the approach. Since the $\hat \phi_i$ form a complete basis, the solution to Eq. (\ref{vectoreq}) is simply
\begin{equation}
\vec \phi = \sum c_i \hat \phi_i e^{\lambda_i x}.
\label{eq:solution}
\end{equation}
The coefficients $c_i$ are determined by the initial ($x = 0$) neutrino flux. For astrophysical neutrinos, this is implemented as an isotropic power law: $\vec \phi_{0} \propto  \vec E^{-\gamma}$, $\tau_0 = 0$, though arbitrary forms of $\vec \phi_{0}$ can be used.

Once the values of ($\lambda_i, \phi_i, c_i$) are determined, Eq. \eqref{eq:solution} can be used to instantly determine the neutrino flux for any zenith angle (which maps to a single column density via Eq. \eqref{eq:coldens})
For electron and muon neutrinos, the components of $C_{ij} = f(E_i,\tilde E_j)$ correspond to the integrand (including the differential element\footnote{In practice one uses a logarithmically spaced energy grid; the differential element then becomes $d\tilde E \rightarrow \tilde E_j ( \log E_{j+1} - \log E_j)$.}) of the last term in Eq. \eqref{eq:transpN}. For tau neutrinos, $C_{ij}$ is the sum of the NC integrand in Eq. \eqref{eq:transpN} and the integrand in Eq. \eqref{eq:taudec}, again including the differential element. The integral over the inelasticity $y$ can be quite time-consuming. In this approach, it only needs to be done once per particle physics model (and PDF set). For the standard model, it can thus be pre-computed and tabulated, so that Eq. \eqref{eq:solution} may be calculated on the fly.

For secondary electron and muon neutrino production from $\tau^\pm$ decay, the approach must be slightly modified as the $\nu_{\ell = e, \mu}$ equations must each be solved simultaneously with the $\nu_\tau$ equation.  This can be written analogously to Eq. \eqref{genericeq}
\begin{equation}
\frac{d}{dx} \left( \begin{array}{c}  \vec \phi_{\nu_\ell} \\  \vec \phi_{\nu_\tau} \end{array} \right) = \left( \begin{array}{c c} - \mathrm{diag} (\sigma) + C_\ell & C_{\tau\rightarrow \ell} \\ 0 &  - \mathrm{diag} (\sigma) + C_\tau \end{array} \right)  \left( \begin{array}{c} \vec \phi_{\nu_\ell} \\  \vec \phi_{\nu_\tau} \end{array} \right) .
\label{eq:combo}
\end{equation}
The $C_{\ell,\tau}$ matrices are the same as in Eq. \eqref{genericeq}; the matrix $C_{\tau \rightarrow \ell}$ is identical to the tau regeneration integrand in Eq. \eqref{eq:taudec}, but using the distribution $dn/dz$ of electron and muon daughter neutrinos.
As in the case of tau regeneration, this matrix only needs to be computed once for a given particle physics model. The solution of Eq. \eqref{eq:combo} is then obtained as before, through the matrix eigenvalues and eigenvectors. 

The method in Eq. \eqref{eq:combo} can in principle be used to couple an arbitrarily large number $M$ of fields, as long as the resulting $NM\times NM$ matrices can be dealt with. It should also be noted that if the on-the-spot approximation of Eq. \eqref{eq:taudec} is insufficient (for example, when energies $\gg$ PeV are required), then the tau cascade equation can also be simultaneously solved as in Eq. \eqref{eq:combo}.

The solution found here is exact in the limit $N \rightarrow \infty$, $E_{N} \rightarrow \infty$, and $E_{i+1} - E_i \rightarrow 0$. In practice, the accuracy is ensured by a grid spacing that is smaller than the scale of any changes in the flux or cross sections, and the requirement that $E_{N}$ be sufficiently larger than the largest energy of interest. This requirement is easy to satisfy for steep power laws like those favored by IceCube data, but breaks down as the incoming neutrino spectrum approaches the unphysical case $\gamma \rightarrow 0$. We have checked that our solution agrees with the publicly available nuSQuIDS package \cite{nusquids}. As a basic benchmarking comparison, obtaining the eigenvector decomposition for a single flavor, including tau-induced secondaries, took 0.11 seconds on a 2.9 GHz laptop CPU. Each subsequent attenuation evaluation took approximately 0.006 seconds. In comparison, a single evaluation of nuSQuIDS for the same energy resolution took 0.2 seconds on a similar machine, once cross section tables had been read (3.5 seconds).  In contrast, the standard Monte Carlo code ANIS \cite{Gazizov:2004va} quotes 300 seconds to generate 10$^4$ samples on a 1 GHz CPU.

\section{Systematic errors on neutrino attenuation}
\label{sec:examples}
We now turn to results of the method described above, as implemented in the \nufate code. Here, we have used the STW105 reference Earth model~\cite{Kutkowski:2008, Trabant:2012}. 
Unless otherwise specified, the neutrino-nucleon cross sections employed in Eq. \eqref{eq:transpN} use the CT10nlo \cite{Nadolsky:2008zw} PDFs, implemented with LHAPDF6 \cite{Buckley:2014ana}. The antineutrino-electron CC cross section is implemented as described in Appendix A of \cite{Palomares-Ruiz:2015mka}.

\begin{figure}
\includegraphics[clip,width=0.5\textwidth]{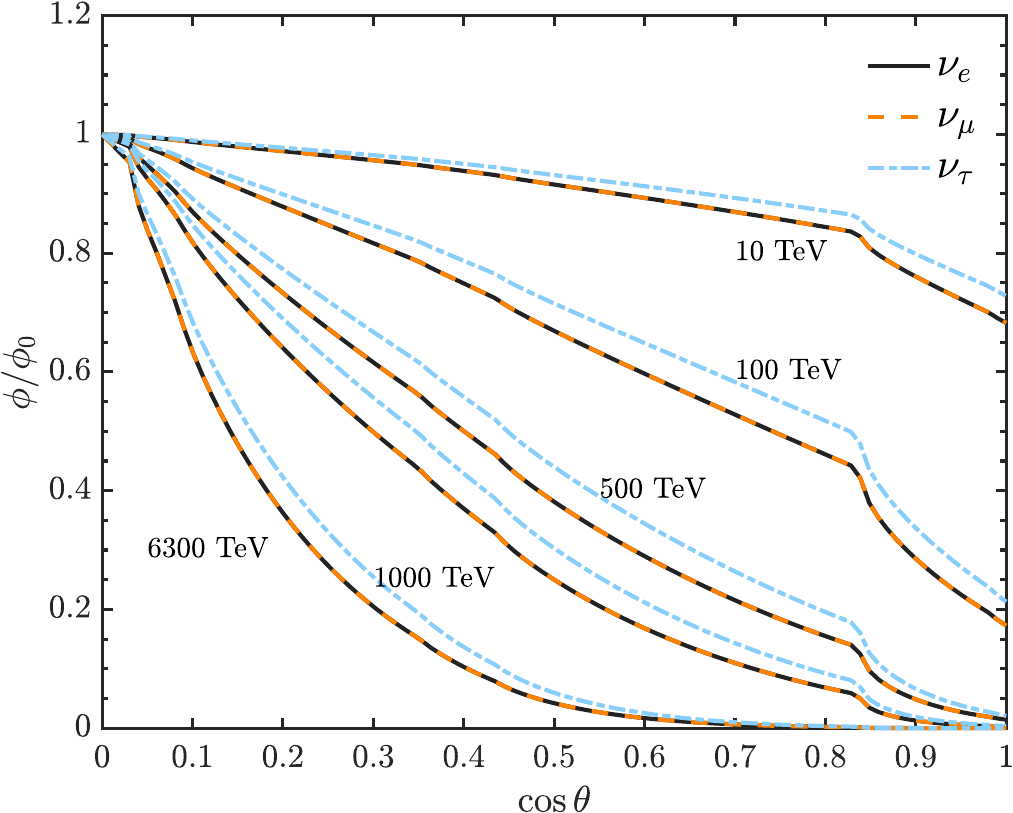}\includegraphics[clip,width=0.5\textwidth]{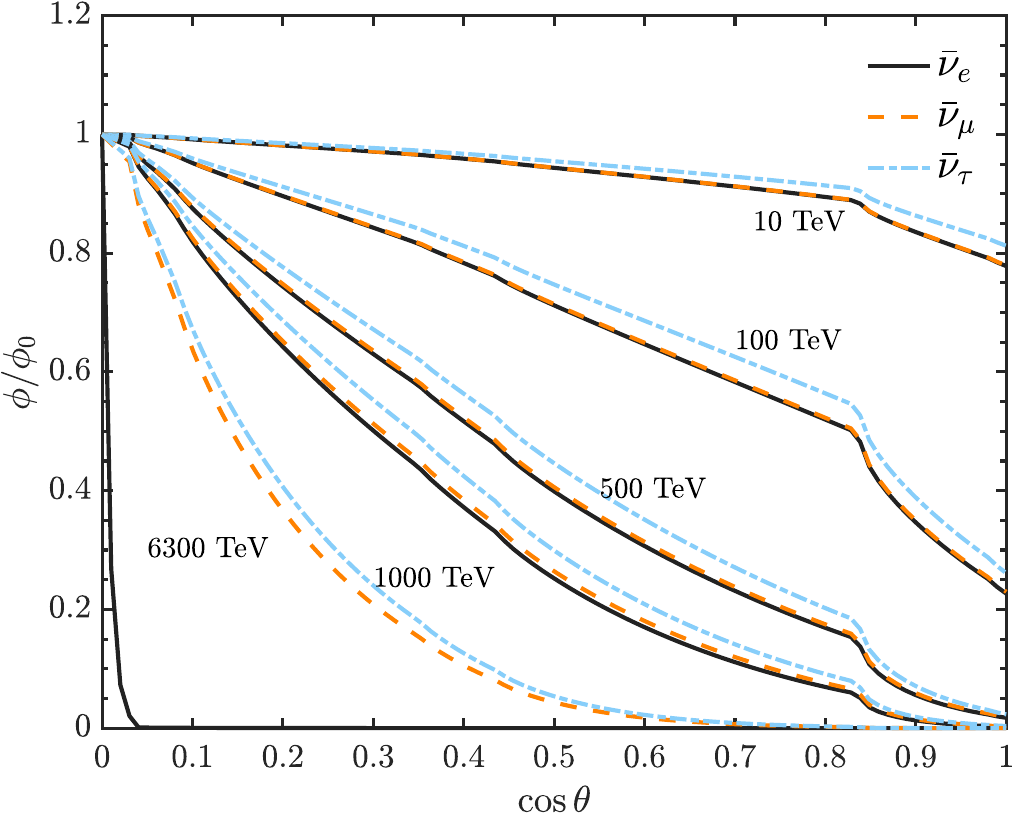} 
\caption{Attenuated neutrino flux as a function of $\cos \theta$, the cosine of the nadir angle, for five different neutrino energies, assuming an isotropic (incoming) neutrino spectrum $\propto E^{-2.5}$. Left: neutrinos; right: antineutrinos.} \label{fig:zenith}
\end{figure}

Fig. \ref{fig:avgfluxes} shows the angle-averaged attenuation for upgoing high-energy neutrinos, under three different hypotheses for the incoming spectrum. As expected, the shape of the initial isotropic flux has a large impact on the measured spectrum. This is especially important for the tau neutrino flux, for which the strong regeneration can be seen in the left-hand panel. In Fig. \ref{fig:zenith}, we plot the (un-averaged)  attenuation as a function of nadir angle (where $\cos\theta= 1$ represents upgoing neutrinos), for several different values of the neutrino energies, assuming an incoming $E^{-2.5}$ neutrino flux. The effect of the Earth's structure can clearly be seen as 'kinks' in the angular distribution. 
\begin{figure}
\includegraphics[trim={0 0 0cm  0},clip,width=0.5\textwidth]{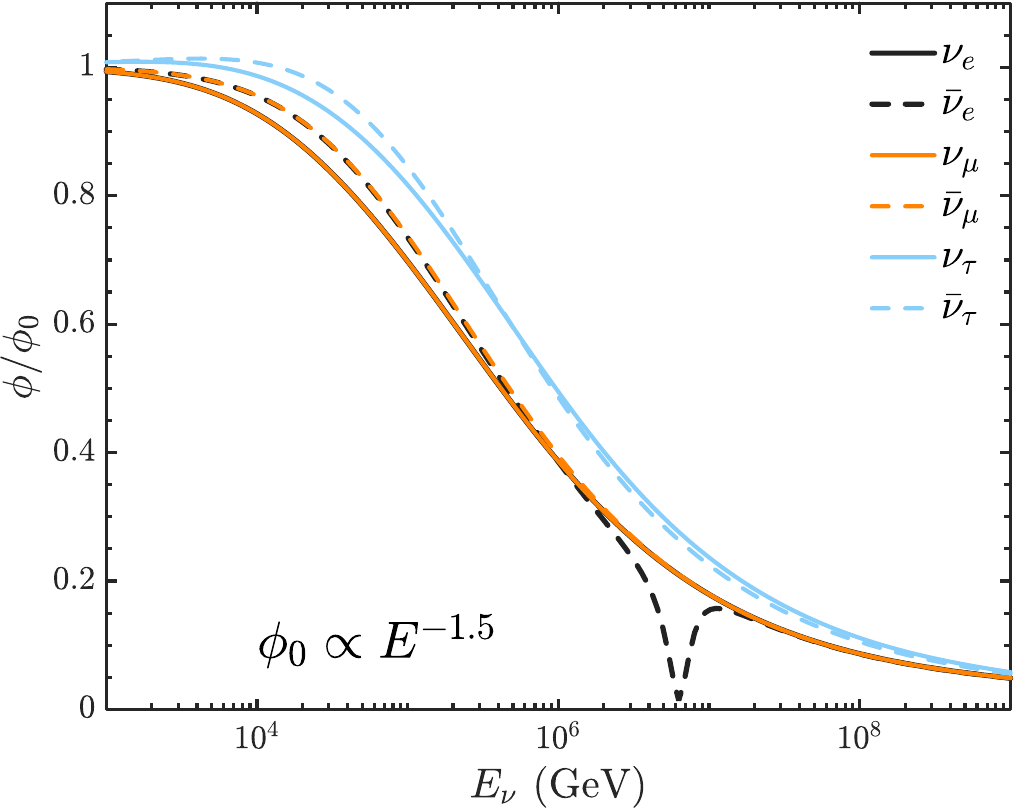}\includegraphics[trim={0 0 0cm 0},clip,width=0.5\textwidth]{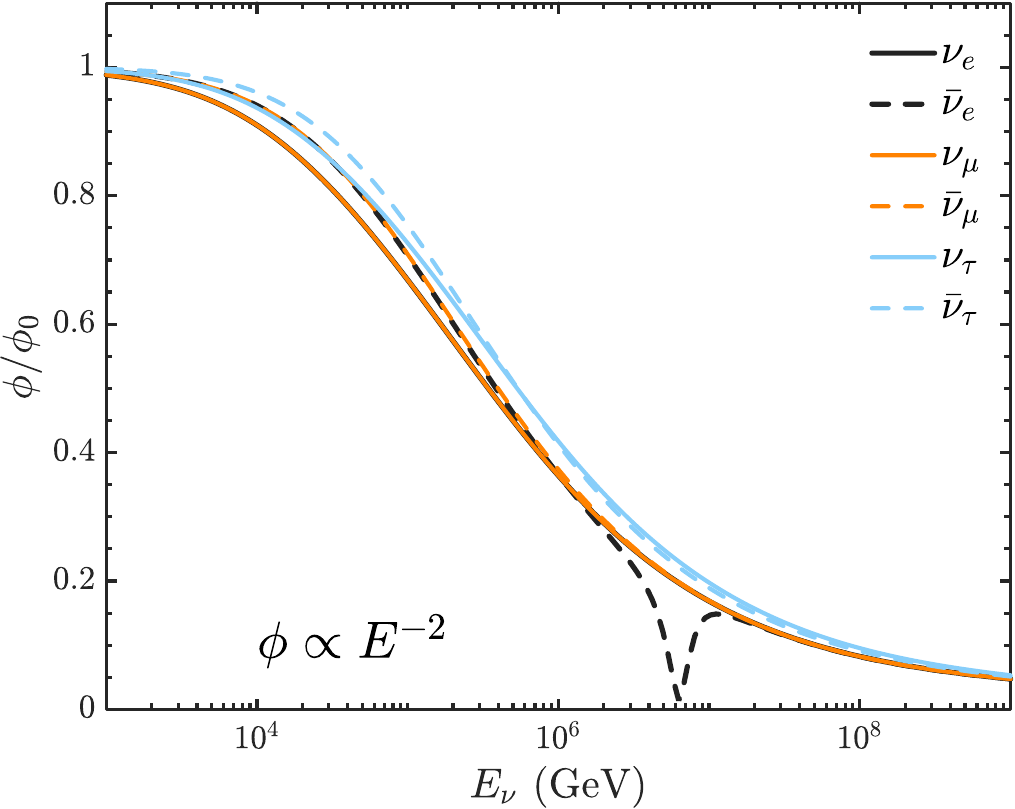} \\\includegraphics[trim={0 0 0cm  0},clip,width=0.5\textwidth]{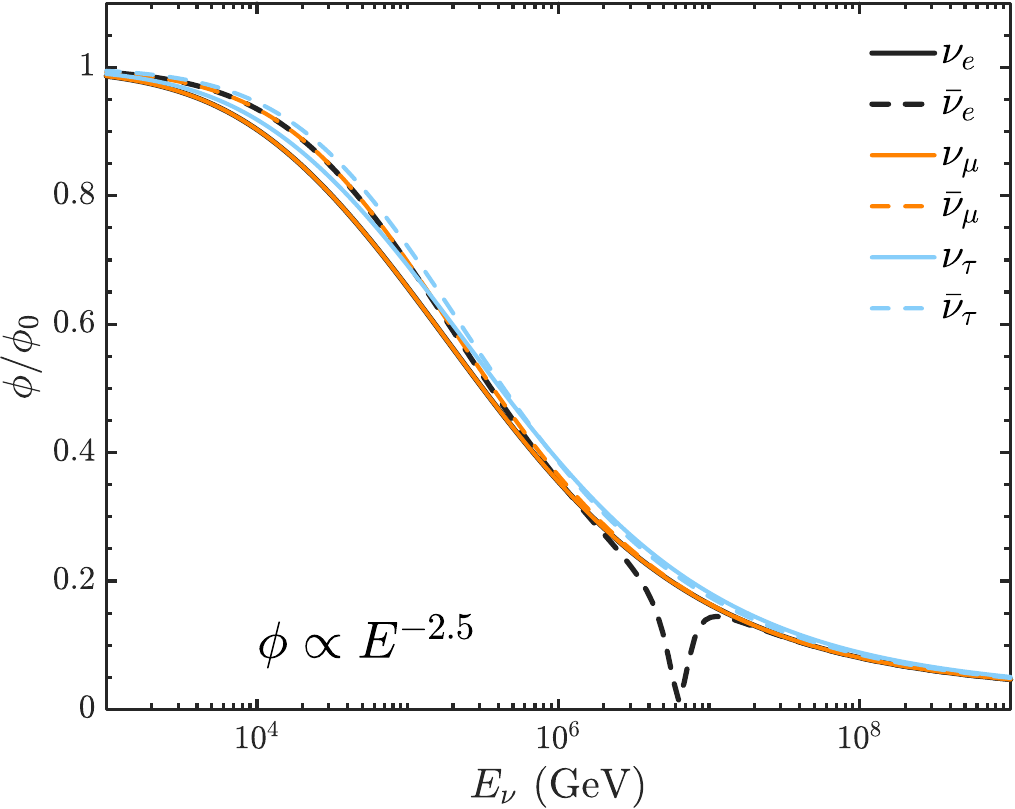}\includegraphics[trim={0 0 0cm  0},clip,width=0.5\textwidth]{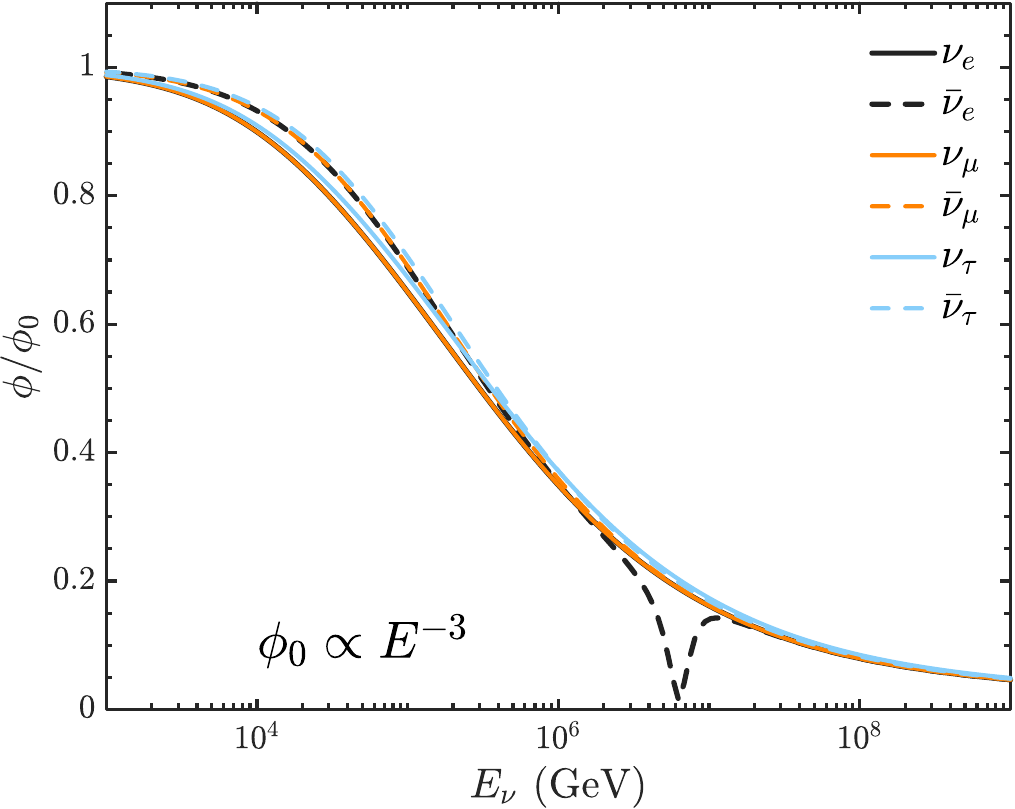}
\caption{Angle-averaged attenuation for upgoing (zenith $\geq$ 90$^\circ$) neutrinos, for four different initial isotropic power laws. Enhanced attenuation due to the Glashow resonance can be seen for $\bar \nu_e$ at 6.3 PeV. The tau (blue) regeneration is particularly sensitive to the power law, since it depends on higher-energy tau neutrinos to source lower-energy ones. } \label{fig:avgfluxes}
\end{figure}

\begin{figure}
\includegraphics[clip,width=.5\textwidth]{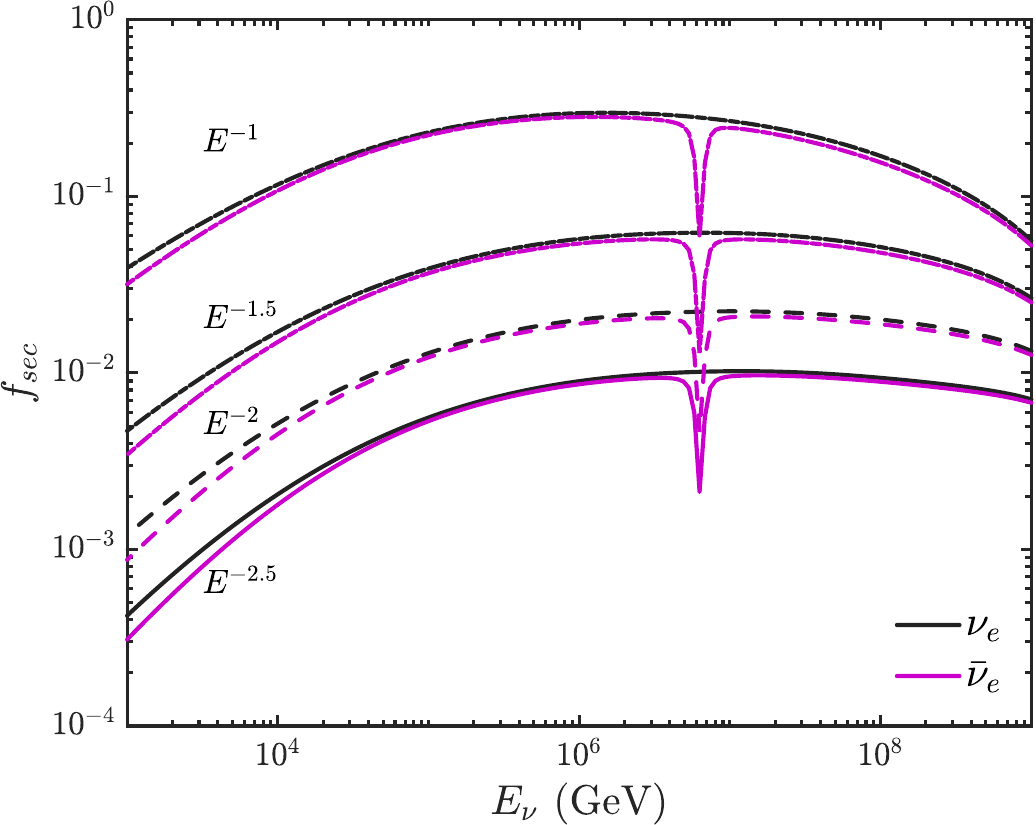}\includegraphics[clip,width=0.5\textwidth]{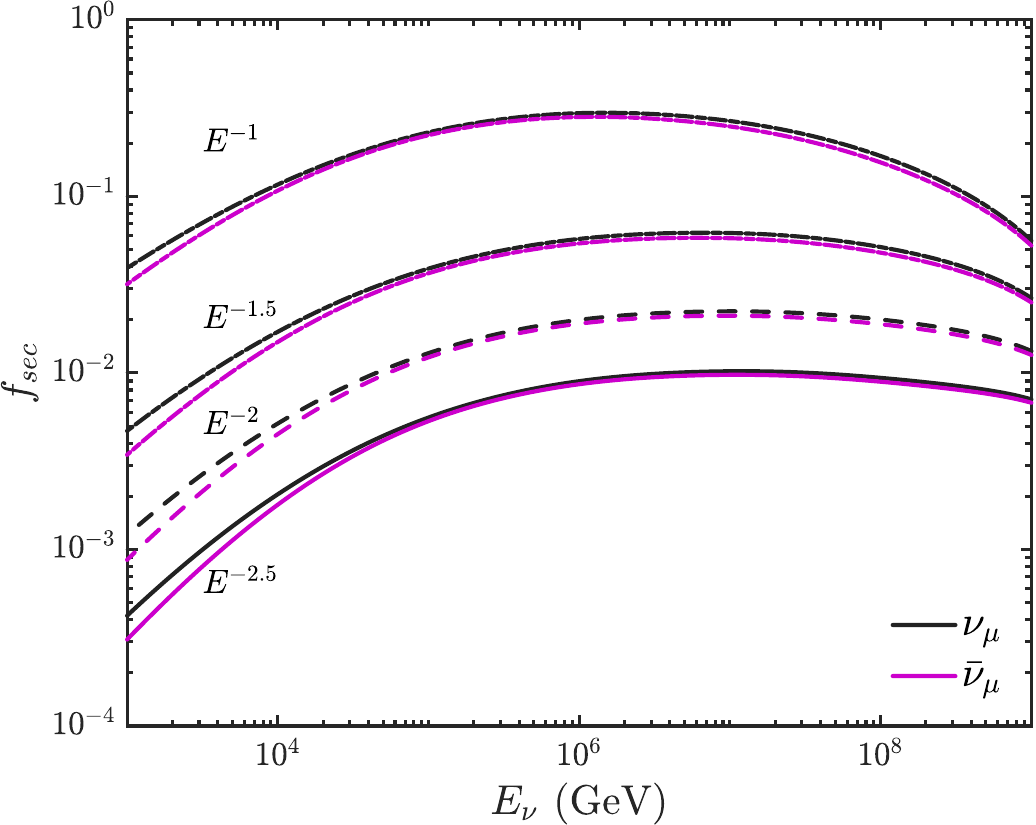}
\caption{Fractional contribution $f_{sec}$ (see Eq. \eqref{eq:fsec}) of secondary electron (left) and muon (right) neutrinos, with respect to the primary fluxes, as computed with Eq. \eqref{eq:combo}, for power laws from $E^{-1}$ (top) to $E^{-2.5}$. (bottom). $\phi_{\nu_\ell,sec.}+\phi_{\nu_\ell,prim.}$ is the total flux computed with \eqref{eq:combo}, while $\phi_{\nu_\ell,prim.}$ neglects secondary production from tau lepton decay. As in Fig. \ref{fig:avgfluxes}, attenuation is averaged over zenith angles $\geq$ 90 degrees.}
\label{fig:secondaries}
\end{figure}

Fig. \ref{fig:secondaries} illustrates the magnitude of secondary electron and muon neutrinos produced with Eq. \eqref{eq:combo}, in comparison with the case where such regeneration is neglected, as in the basic case in Eq. \eqref{vectoreq}
\begin{equation}
f_{sec} = \frac{\phi_{\nu_\ell, sec.}}{\phi_{\nu_\ell, prim.}}.
\label{eq:fsec}
\end{equation}
As is well known, this effect becomes very important for harder spectra, for which energetic tau leptons are copiously produced in the Earth.  

In the following sections, we explore the impact of different external parameters on the attenuation, using the quantity ${\delta Att(E_\nu)}/{Att(E_\nu)}$,where $Att(E_\nu)$ represents the zenith-averaged factor by which the flux is attenuated by the Earth
\begin{equation}
Att(E_\nu) \equiv \frac{ \int_{-1}^0 \frac{d\phi}{dE} d\cos(z)}{{d\phi_0}/{dE}}
\end{equation}
and $\delta Att(E_\nu)$ represents the absolute error on that factor. ${\delta Att(E_\nu)}/{Att(E_\nu)}$ thus represents the relative error on the attenuation, and is the relevant quantity to compute, in that it is equal to the resulting relative error in the upgoing neutrino flux seen at experiments. 

\subsection{Incoming neutrino spectrum}
\begin{figure}
\includegraphics[clip,width=.5\textwidth]{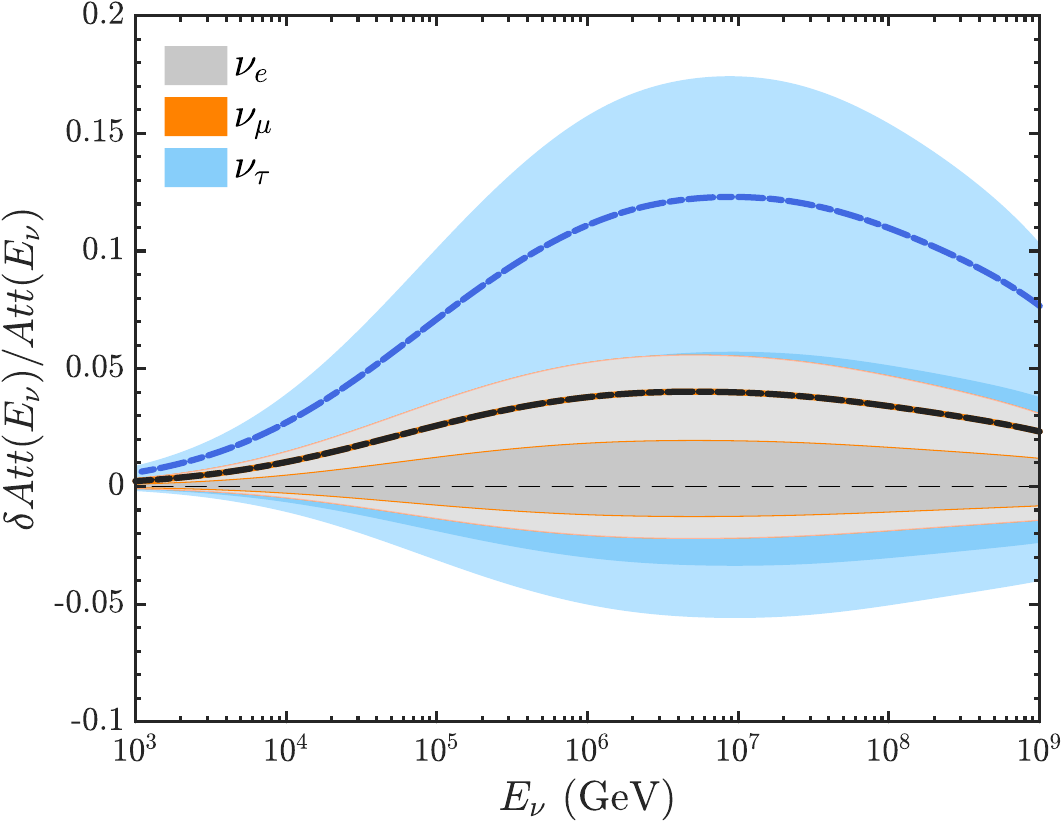}\includegraphics[clip,width=0.5\textwidth]{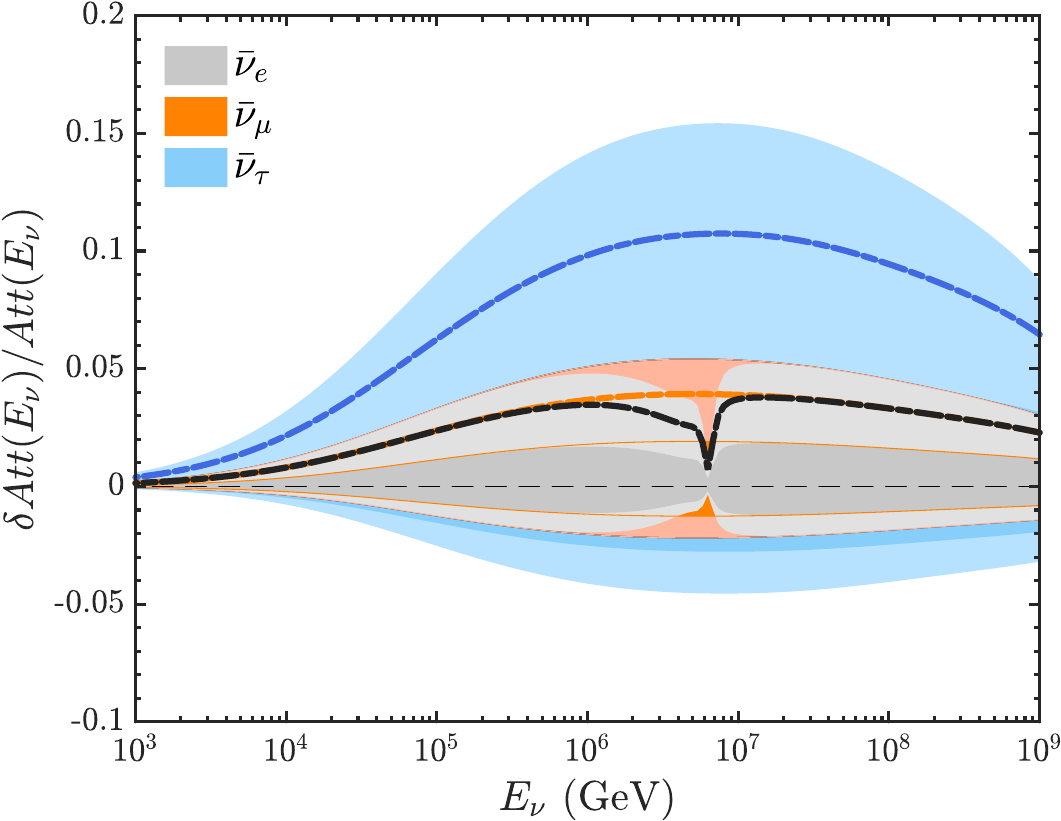}
\caption{One (dark) and two (light) sigma error in the attenuation for upgoing neutrino flux due to the uncertainty in the incoming neutrino spectrum power law: $d\phi/dE \propto E^{-\gamma}$, $\gamma = 2.83 \pm 0.50$ \cite{Vincent:2016nut}. $\delta Att = 0$ corresponds to the best fit $\gamma = 2.83$ case. This propagates to an error in the effective area $A_{eff}(E_\nu)$, as defined e.g. in IceCube publications: $A_{eff}(E_\nu) \propto Att(E_\nu)$. The thick dash-dotted lines correspond to the $E^{-2}$ case used by IceCube. This highlights the importance of separating out the attenuation of different flavors from the detector sensitivity to CC and NC processes, as done in Refs \cite{Palomares-Ruiz:2015mka,Vincent:2016nut}. The asymmetry is due to the effect of secondary neutrino production from tau lepton decay in harder incoming spectra. Note also that the orange (muon) regions are not visible since they lie directly behind the electron (gray) regions. }
\label{fig:indexfig}
\end{figure}

The zenith ($z$) dependence as well as part of the energy dependence of the effective areas provided by IceCube \cite{Aartsen:2013jdh} are due to the direction and energy dependence of attenuation in the Earth. However, these are computed for a specific isotropic spectral index. A full analysis of the HESE data requires a consistent computation of the effective area for a given incoming neutrino flux $\propto E_\nu^{-\gamma}$, as performed, e.g., in \cite{Palomares-Ruiz:2015mka,Vincent:2016nut}. Fig.~\ref{fig:indexfig} shows the relative difference in attenuation rates for upgoing high-energy neutrinos due to the variation of the spectral index within values allowed from self-consistent analysis of the upgoing data only \cite{Vincent:2016nut}: $\gamma = 2.83 \pm 0.50$. This can be as large as 20\% for tau neutrinos, due to the importance of regeneration at lower spectral indices, and highlights the importance of separating out attenuation from effective area in such analyses, as explained, e.g., in \cite{Palomares-Ruiz:2015mka}, rather than including attenuation inside the effective area. 

\subsection{Uncertainties from Earth density model}
\label{sec:rho}
The same method can be used to evaluate effects of uncertainties in the density profile of Earth on the attenuation of the neutrino flux in the Earth. The Reference Earth Model (REM) employed here is STW105 \cite{Kutkowski:2008, Trabant:2012}, which can be parametrized via second order polynomials, separated into eight components corresponding to the different layers in the Earth. We allow for local density variations of up to 10\% (these could be even higher, see  \cite{chambat2001mean,Masters2003159}), although several other constraints need to be satisfied. These include the total mass $M$, moment of inertia $I$, and the requirement that density is strictly decreasing  \cite{CarlosThesis,Jones:2015bya}. To model this variability, we create 500 models with the following procedure, repeated a random number (up to 6) of times:\footnote{The method described in \cite{CarlosThesis,Jones:2015bya} also creates valid models.}
\begin{enumerate}
\item Randomly pick one of the eight segments in the REM. 
\item Randomly rescale its height and slope, under the conditions that a) there is no deviation of more than 10\% at any point, b) the slope remains negative, and c) that the density remains lower than the previous segment and higher than the next. We additionally require the derivative to vanish at $r = 0$.
\item Randomly pick two segments and rescale them so that $\delta M = \delta I = 0$, subject to the above conditions. If this is not possible, keep picking new segments until these requirements are satisfied.
\item  Repeat.
\end{enumerate}
\begin{figure}
\begin{tabular}{c c}
\includegraphics[clip,width=.485\textwidth]{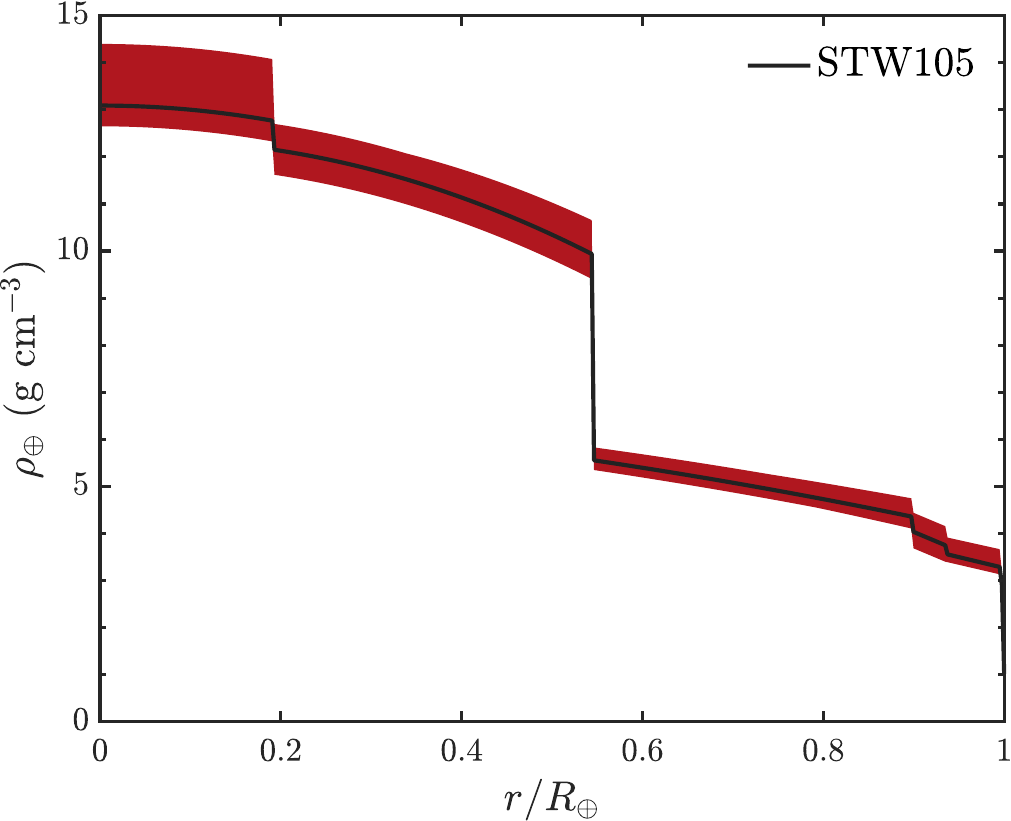} &\includegraphics[clip,width=0.5\textwidth]{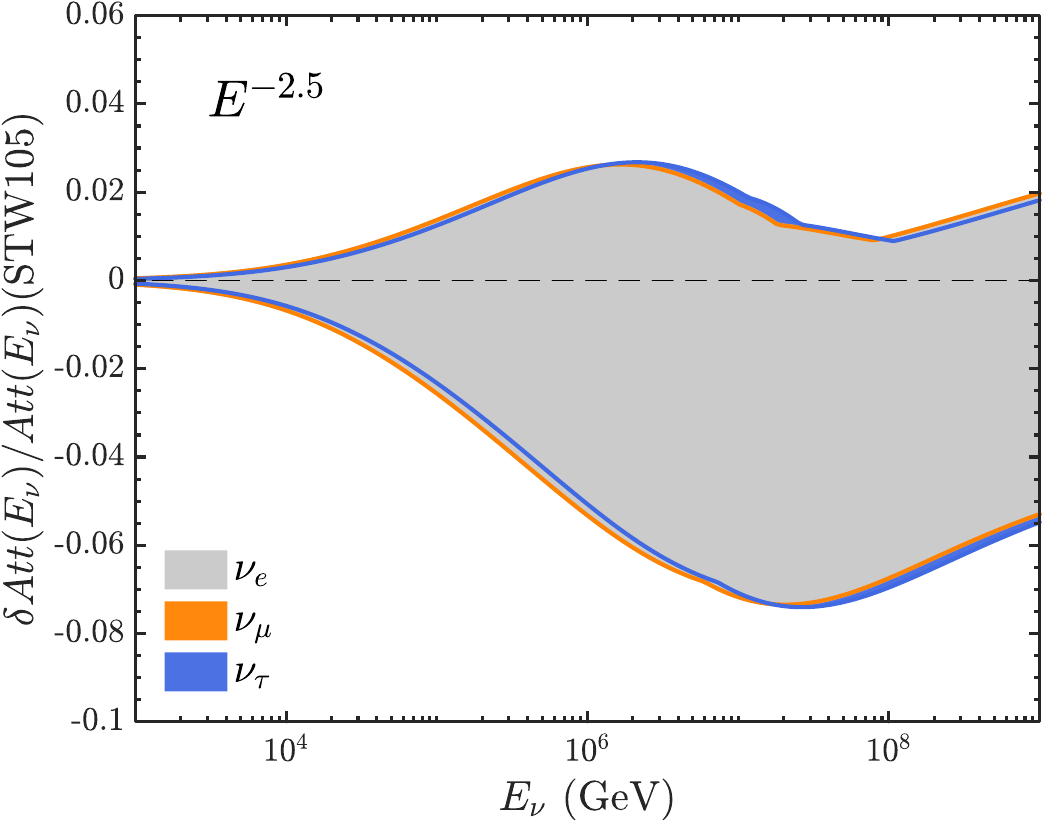} \\
\includegraphics[clip,width=.5\textwidth]{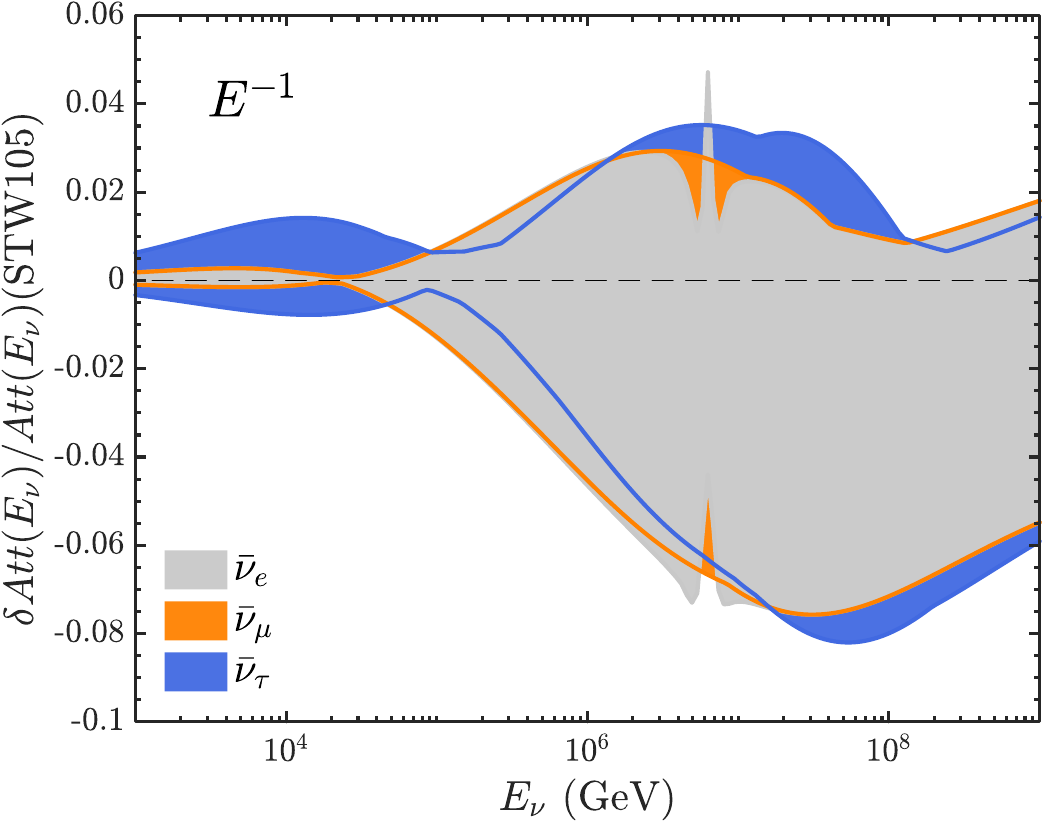} &\includegraphics[clip,width=0.5\textwidth]{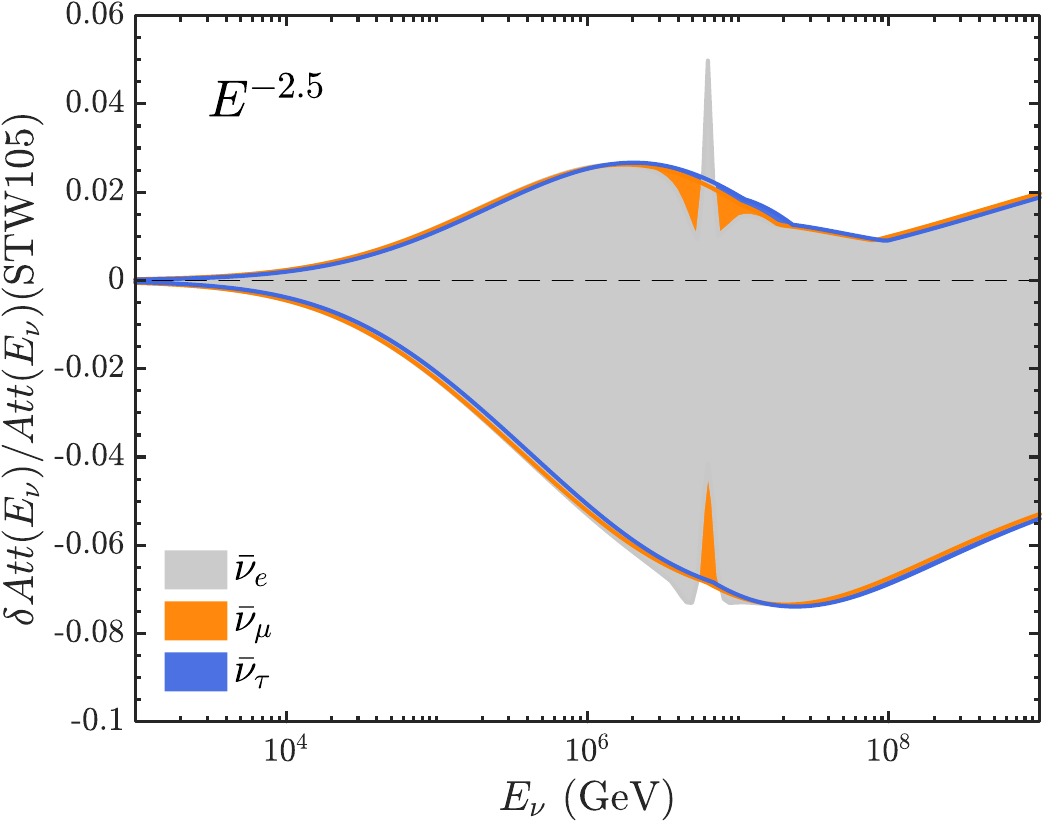} 
\end{tabular}
\caption{Fractional change in the attenuation for upgoing neutrino flux due to uncertainties in the Reference Earth Model (REM), which propagates to an error in the effective area $A_{eff}(E_\nu)$, as defined e.g. in IceCube publications: $A_{eff}(E_\nu) \propto Att(E_\nu)$. Top-left: envelope of models allowed by conditions enumerated in Sec \ref{sec:rho}. Reference model in black is STW105 \cite{Kutkowski:2008, Trabant:2012}. Right: relative variation in attenuation due to differences in these models for (top) neutrinos and (bottom) antineutrinos for an incoming flux $\propto E_\nu^{-2.5}$. The shape and magnitude of these changes is largely independent of the incoming neutrino spectral index, except for very hard spectra where secondary production from $\tau^{\pm}$ decay becomes very important. This is shown in the bottom-left panel for $\phi_0 \propto E^{-1}$. } \label{fig:Earthdens}
\end{figure}
The resulting range of models is shown in the upper-left panel of Fig. \ref{fig:Earthdens}, keeping in mind that models within the envelope shown must also satisfy the conditions enumerated above. The right-hand panels of Fig.~\ref{fig:Earthdens} show the resulting envelope of neutrino flux attenuation for the zenith-averaged upgoing neutrino flux in the case of an $E^{-2.5}$ incoming power law flux, showing that uncertainty due to the Earth's composition can be as large as 8\%. Varying the power law has little qualitative effect, unless the spectrum is allowed to be very hard. The lower-left panel shows the same effect, but for antineutrinos with an $E^{-1}$ flux. The contribution from secondary regeneration can readily be seen. 

\subsection{Uncertainties on neutrino cross sections}
\label{sec:xsecs}
Neutrino attenuation through the Earth depends on the neutrino-nucleon cross section,  dominated by deep inelastic scattering (DIS) processes. The DIS cross section uncertainties have been summarized in \cite{Formaggio:2013kya, CooperSarkar:2011pa}. These are due to differences in the parton distribution functions (PDFs) which specify the quark composition $q(x,Q^2)$ (and antiquark, $\bar q(x,Q^2)$) of the nucleons as a function of the kinematic variables. In order to estimate the possible impact of PDF uncertainties, we evaluate the results of \nufate propagation using three representative PDF sets which use different methods: CT10nlo \cite{Nadolsky:2008zw}, HERAPDF 1.5 \cite{Aaron:2009aa}, and NNPDF 23 \cite{Ball:2012cx}. This choice is motivated by the fact that each PDF collaboration uses a different systematic approach to construct the PDF: NNPDF uses a neural network to select functions to fit the data, HERAPDF uses only the information collected by experiments at HERA that allows for robust systematic treatment, and CT10nlo combines different experiments while using a more traditional phenomenologically motivated approach for the base functions. We use the central values, along with the one sigma uncertainty band given in each PDF set. We use the following procedure: since ${d\sigma}/{dxdy} \propto \sum_i F_i$ where the structure functions $F_i = \sum_j A^i_j q_j(x,Q^2) + B^i_j \bar q_j(x,Q^2)$, we calculate the Jacobian, $(J_F)_{ij} = A^i_j + B^i_j$. We then use the reported covariance matrix on $q_j(Q,x)$ provided with each PDF. This defines the confidence limits in $F_i$, which we then propagate to the differential cross sections and which are in turn implemented into \nufate. This cross section calculation is analogous to the one shown in \cite{Arguelles:2015wba}, and the error propagation procedure will be published in \cite{Arguelles:2017inprep}. 

\begin{figure}
\begin{tabular}{cc}
\includegraphics[trim={1cm .1cm 1cm 1.05cm},width=0.5\textwidth]{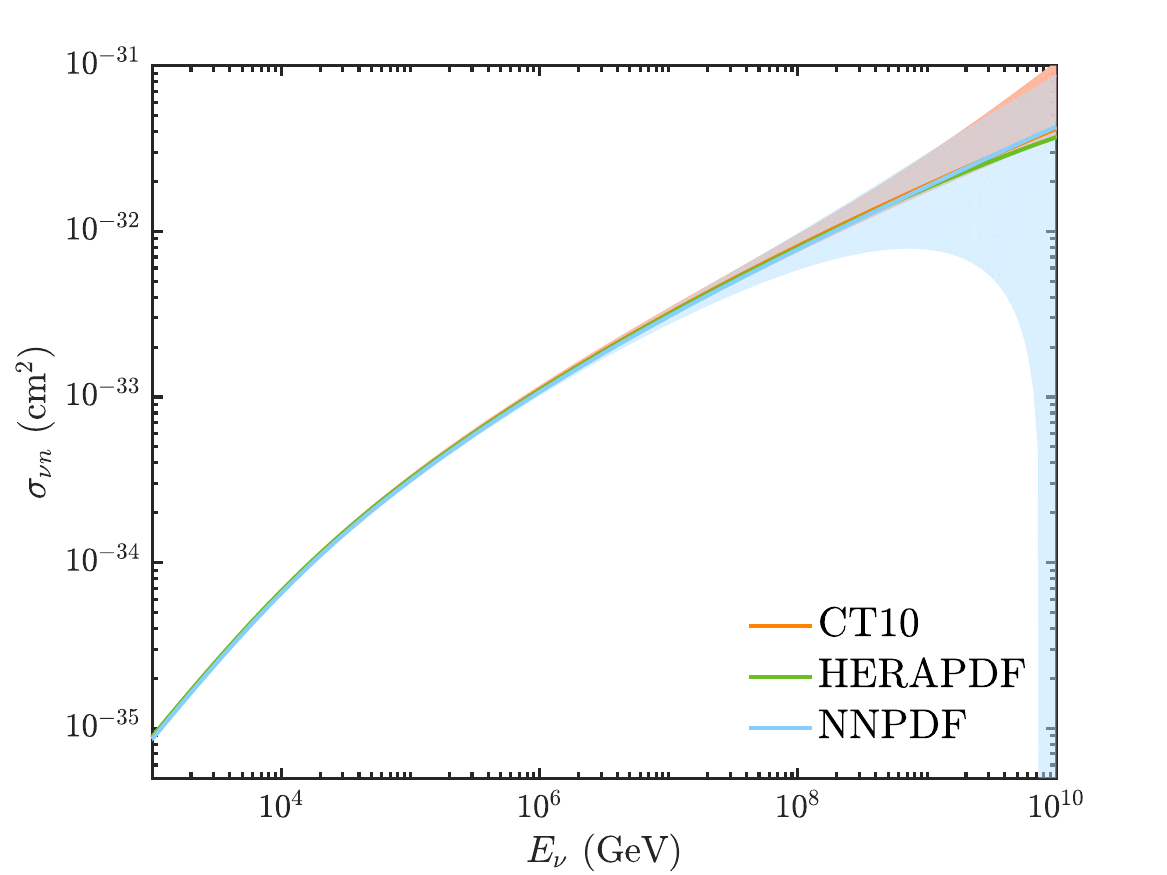}& \includegraphics[width=0.5\textwidth]{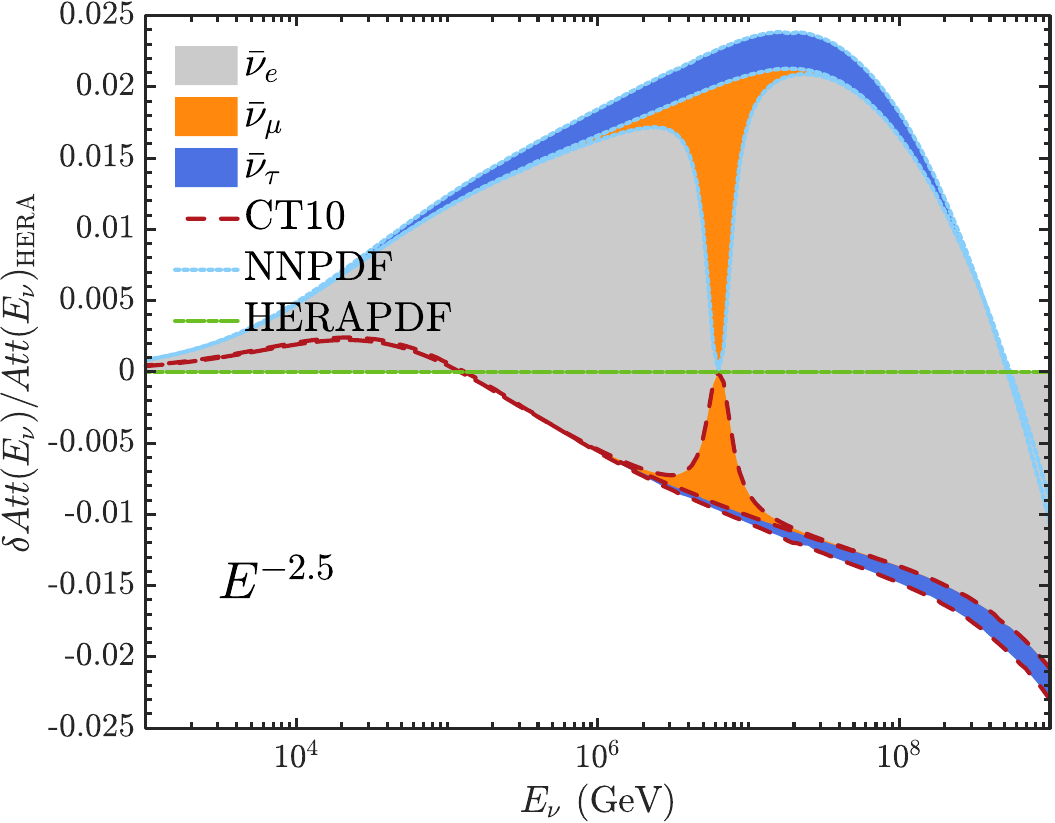}
\end{tabular}
\caption{Left: the neutrino-nucleon cross sections using three different PDF sets and their their $3\sigma$ errors (shaded). Right: the relative deviation with respect to HERAPDF of the PDF sets under consideration (HERA is chosen as a reference simply because its values mainly lie between the other sets). The shaded regions show the maximum deviation for each antineutrino flavor, and the lines show the flavor-dependent deviation for each of the PDF sets. Effects of uncertainties are not included; these are shown in Fig. \ref{fig:attenuation_uncertainty}.}
\label{fig:pdf_sets}
\end{figure}

The first panel of Figure \ref{fig:pdf_sets} shows the total neutrino-nucleon cross section using each of the PDF sets, along with their 3$\sigma$ CL bands. The right-hand panel shows the relative differences between attenuation rates in the upgoing flux, using the central values of the three PDF sets under consideration. The agreement is relatively good: below neutrino energies of 10 PeV, the total range is less than 4\% of the average flux. 

\begin{figure}
\begin{tabular}{cc}
\includegraphics[width=0.5\textwidth]{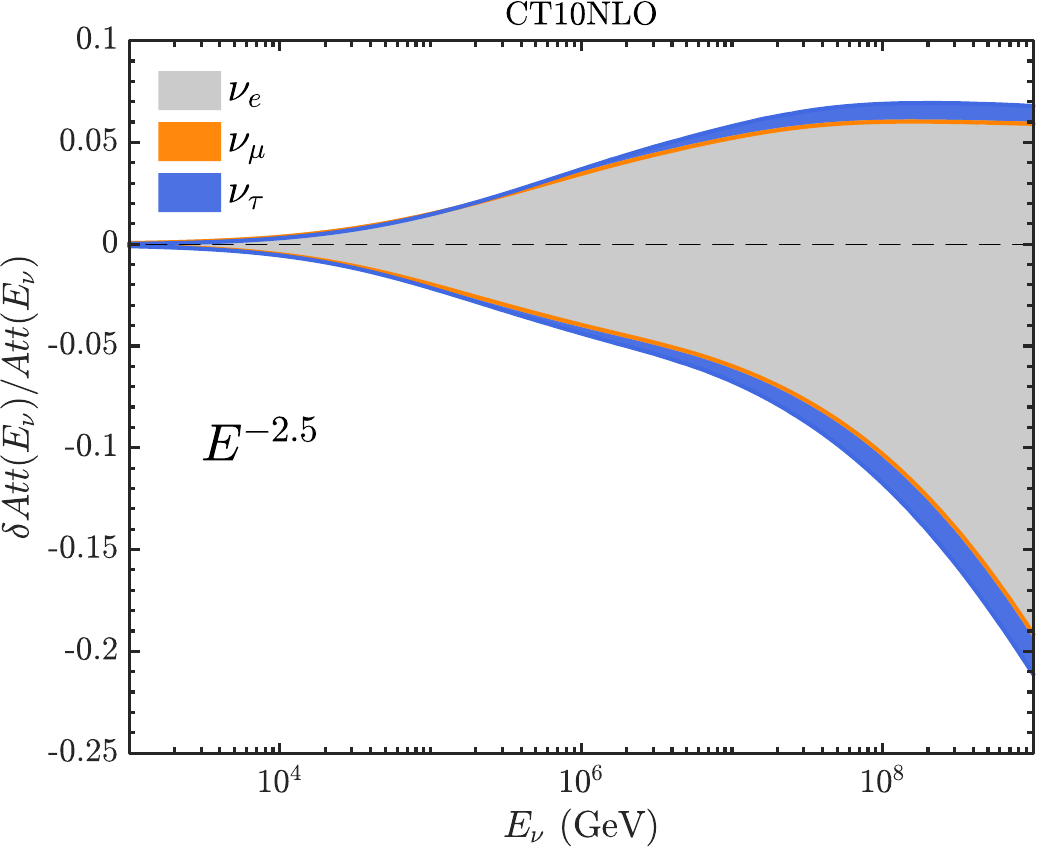}& \includegraphics[width=0.5\textwidth]{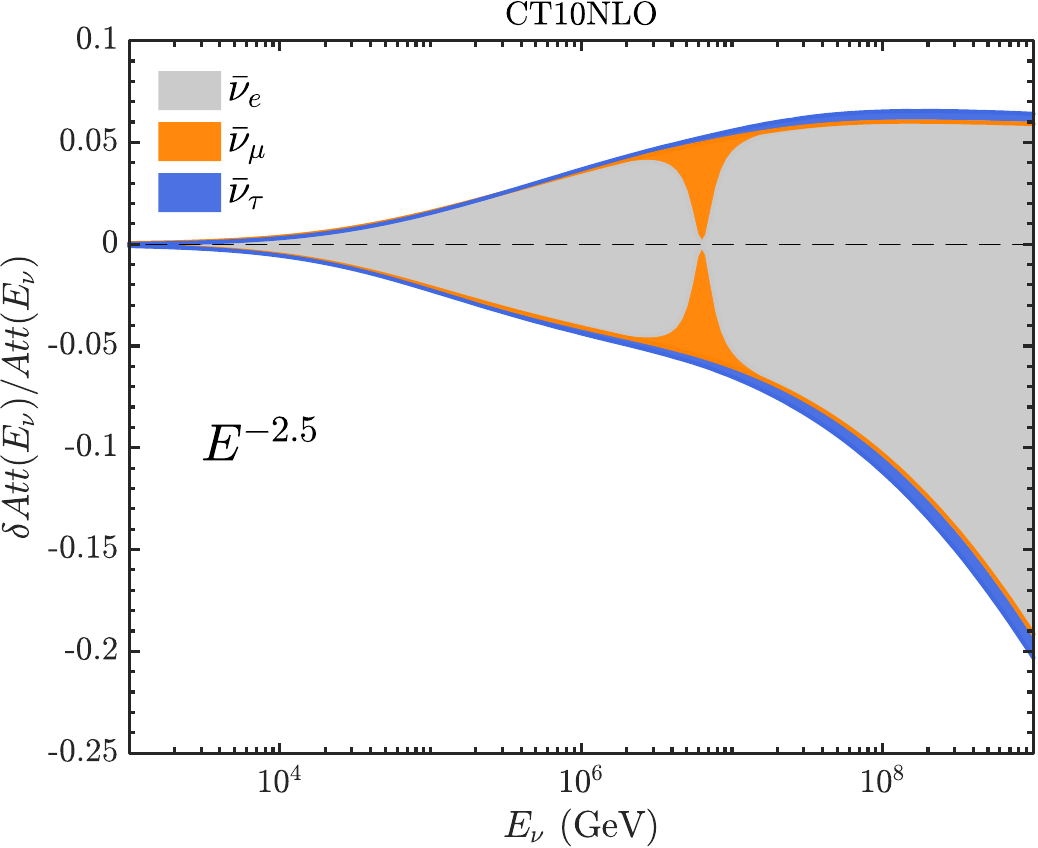}\\
\includegraphics[width=0.5\textwidth]{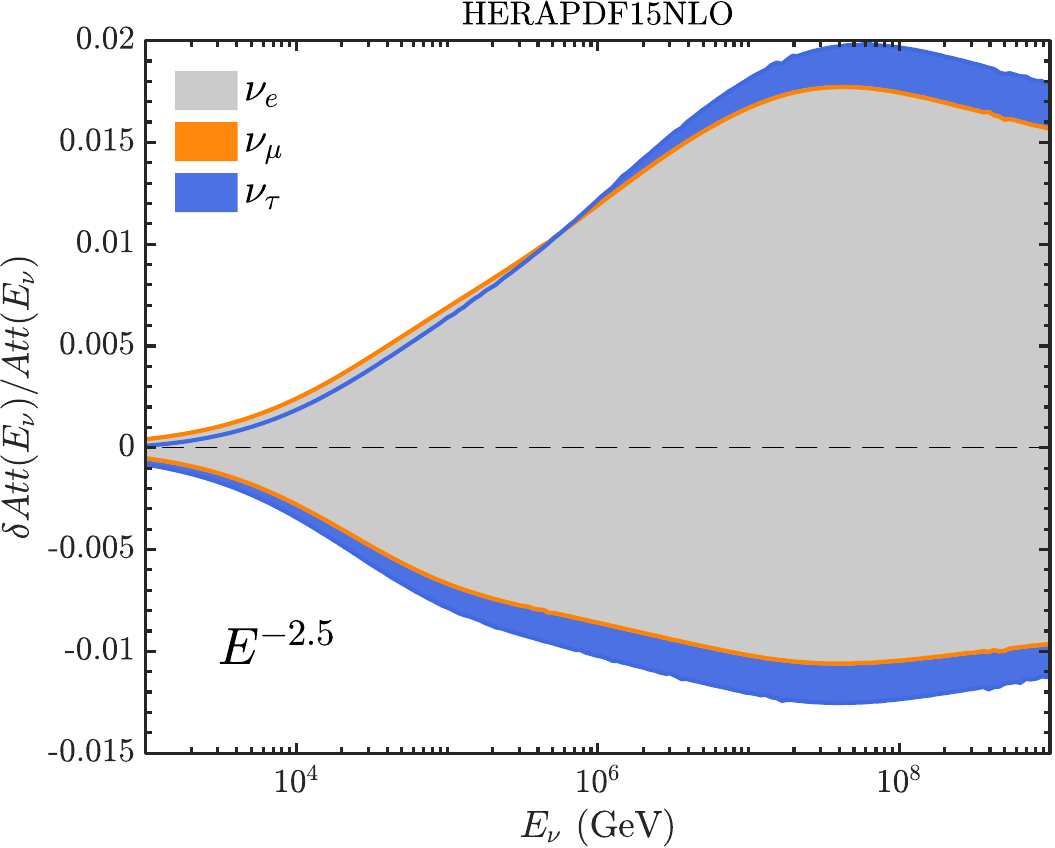}& \includegraphics[width=0.5\textwidth]{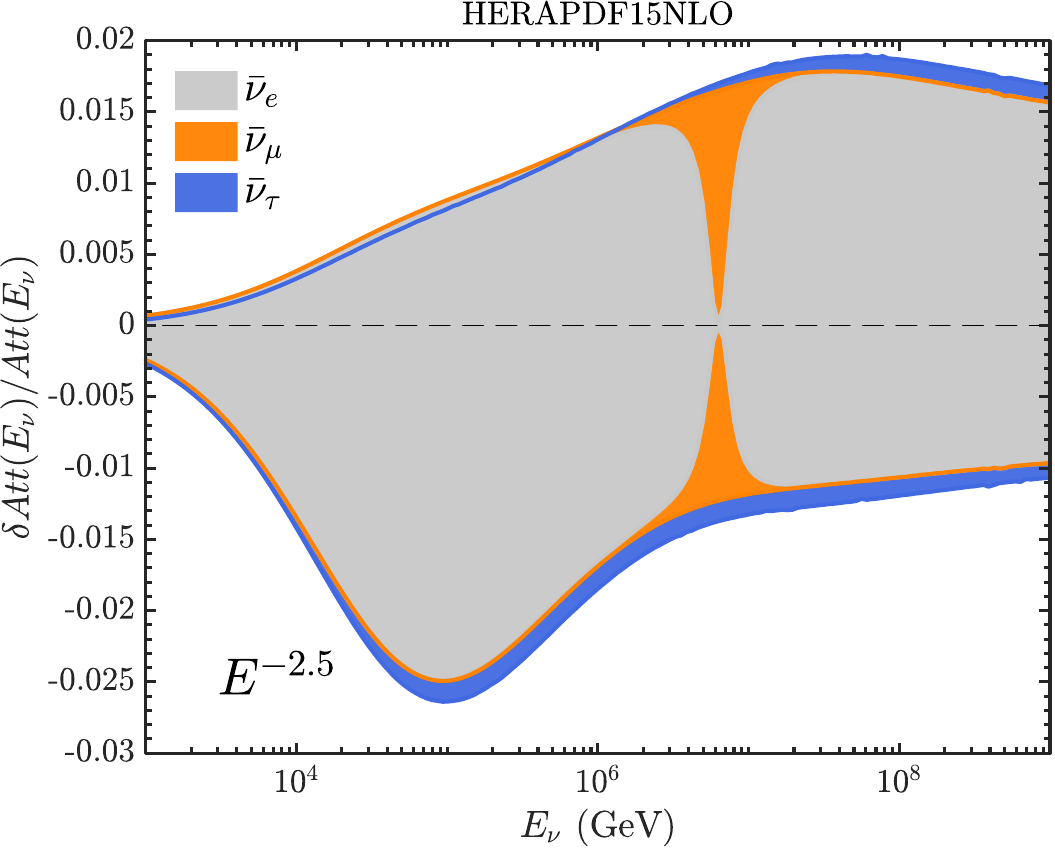}
\end{tabular}
\caption{Relative uncertainty ($3\sigma$) on the attenuation of the neutrino (left) and antineutrino (right) flux through earth, due to the uncertainty on the parton distribution functions used to compute the neutrino-nucleon cross sections. Top: CT10 PDF set; bottom: HERAPDF. We do not show the equivalent plot for NNPDF due to the very large uncertainties at high energies (see Fig. \ref{fig:pdf_sets}).  }
\label{fig:attenuation_uncertainty}
\end{figure}

Fig. \ref{fig:attenuation_uncertainty} shows the propagated $3\sigma$ error on the attenuated flux for two of the PDF sets (CT10, top, and HERAPDF, bottom). These errors are clearly much larger in the case of CT10 (up to 20\% deviation), than in the newer HERAPDF case (at most 2.5\%). We have only shown the errors for an $E^{-2.5}$ astrophysical flux. These are of course slightly larger for harder fluxes. The main impact, as before, is on the tau neutrino flux due to enhanced regeneration from $\tau^\pm$ decay. 

We have not show the NNPDF case due to the very large reported uncertainties at high energies: these are due to the fact that NNPDF is the only set that goes to very low values of the Bjorken $x$ parameter. This is crucial, as the scattering processes that are relevant here are in the very forward (low transferred momentum) regime, where low-$x$ effects -- and their associated uncertainties -- are critical. It is thus likely that the uncertainties that we have shown based on the other two PDF sets are underestimated, as these rise significantly for small values of $x$. Because of this, the addition of data from experiments such as LHCb (see \cite{Gauld:2016kpd}) will be crucial in pinning down high-energy neutrino-nucleon cross sections.



\section{Summary}
We have described a fast method of solving the cascade equation that describes the attenuation of high-energy neutrinos as they propagate through the Earth. We provide as an example a prototypical isotropic astrophysical $Eˆ^{-\gamma}$ flux.  Details of an implementation of this method are given in Appendix \ref{sec:app}, which includes the capability to read in an arbitrary flux from a text file. The \nufate method is fast and efficient, as long as the propagation ODE depends only on the column density of nuclei. When the tau lepton lifetime becomes relevant, or if oscillation e.g. into a sterile state occurs, a full ODE solver such as nuSQuIDS becomes nearly unavoidable.\footnote{In this case the cascade equation can be rewritten $\vec \phi'(x) = C(x)\vec \phi(x)$, so that the eigenvectors of $C$ are no longer constant in $x$.} We have also provided three examples of this method's use in the propagation of systematic errors that can significantly affect attenuation at high energies: 
\begin{enumerate}
\item We explicitly show that the current error on the incoming neutrino spectral index results in an uncertainty of up to 20\% on the attenuated flux, with the best fit yielding a  10\% difference in effective area with respect to the tables provided by the IceCube Collaboration.

\item Uncertainties on the Earth model can lead to changes as large as 10\% with respect to computations using a Reference Earth Model. 

\item By considering three different parton distribution functions, we demonstrate how the uncertainties in the deep inelastic scattering cross section measurements could impact the uncertainties in the attenuation of neutrinos passing through the Earth, creating an uncertainty up to the 10\% level  at PeV energies. 
\end{enumerate}
Because the upgoing neutrino event rate is directly proportional to the energy-dependent attenuation factor, these uncertainties propagate to determinations of the astrophysical neutrino flux and spectral index, as well as the isotropy, in addition to any new physics search that depends on zenith angle. As current statistical errors approach O(1) of the observed flux, it is all the more important to constrain these systematics to the percent level. Meanwhile, as these uncertainties grow with energy, their impact on the observation and sensitivity to the GZK neutrino flux are expected to be more dominant.

The above examples highlight the importance of treating attenuation independently from the effective area, as the systematics that govern each term are very different.

\acknowledgements
We thank John Beacom, Rhorry Gauld, Mary Hall Reno, Sergio Palomares-Ruiz and Chris Weaver for useful comments and suggestions. ACV is supported by an Imperial College London Junior Research Fellowship (JRF). CAA is supported by NSF grants No. PHY-1505858 and PHY-1505855. AK was supported in part by NSF under grants ANT-0937462 and PHY-1306958 and by the University of Wisconsin Research Committee with funds granted by the Wisconsin Alumni Research Foundation.

\bibliographystyle{apsrev4-1}
\bibliography{nuFATE}

\appendix
\section{The \nufate code}
\label{sec:app}
The \nufate code is available from this URL: \url{https://github.com/aaronvincent/nuFATE}. The Python version requires only a working installation of Python 2.7 and the standard numpy libraries. \nufate comes with the following Python files:
\begin{description}
\item[earth.py]  computes the optical depth for a given angle. It uses a piecewise polynomial fit to the STW105 model. Any model of the Earth can in principle be substituted.
\item [cascade.py] computes the eigenvectors and eigenvalues required to solve the cascade equations, using 200 $\times$ 200 precomputed arrays located in the \texttt{data/} sub-directory. This includes the effects of attenuation, neutral current down-scattering, as well as tau regeneration. For secondary $\nu_e$ and $\nu_\mu$, use \texttt{cascade\_secs.py}. In practice, \texttt{cascade.py} actually solves the cascade equation for $E^2 \phi$ rather than $\phi$, to avoid having to deal with large flux differences.  The input argument \texttt{gamma} can be either a number $\gamma$, in which case the initial isotropic neutrino spectrum is taken to b
\item [example.py] is a working example of how to call the routines. The neutrino flavor, isotropic spectral index, zenith angle and energy can be specified in the file. From a python prompt it can be executed by calling $$\texttt{>>> import \textbf{example}}$$ If the code runs successfully, you should see something like: $$\texttt{Flux at E  = 100000.0  GeV , zenith =  130.0  degrees will be attenuated by a factor of  0.597606261699}$$
\item [notebook.ipynb] is a python notebook that contains a few more examples and plots. It can be opened and run with iPython or Jupyter notebooks\footnote{See \url{http://jupyter.readthedocs.io/en/latest/}}
\end{description}
Equivalent files are available in Matlab format in the \texttt{matlab/} directory. These have been tested in R2016b. The \texttt{examples.m} script contains two working examples. 

The cross sections, computed with CT10 NLO pdf set, are stored in h5 (high density) format in the \texttt{data/} directory. These include total charged-current and neutral-current neutrino-nucleus scattering, as well as the differential cross sections required for NC interactions and tau decay. The Earth composition is assumed to be isoscalar, such that the cross sections are an average of the neutrino-proton and neutrino-neutron interactions. In the case of $\bar \nu_e$, interactions with electrons are included directly in the \nufate code, since their analytic expressions may be computed on the fly. 
\end{document}